\documentclass[aps,preprint]{revtex4}%
\usepackage{amsfonts}
\usepackage{amsmath}
\usepackage{amssymb}
\usepackage{graphicx}%
\begin{document}
\preprint{ }
\title{\textbf{Effects of quantum statistical pressure and exchange correlation on
the low frequency electromagnetic waves \ in degenerate Fermi-Dirac pair-ion
plasma}}
\author{Zahida Ehsan$^{1}$, M. Shahid$^{2}$, M. A. Rana$^{1}$  A. Mushtaq$^{3},$ A.
Abdikian$^{4}$ and A. Shahbaz$^{5}$ }
\affiliation{$^{1}$Department of Physics, COMSATS Institute of Information Technology,
Lahore 54000, Pakistan}
\affiliation{$^{2}$Department of Physics, Women University of Azad }
\affiliation{$^{3}$ Department of Physics, FBAS, International Islamic University (IIUI),
Islamabad 44000, Pakistan.}
\affiliation{$^{4}$Department of Physics, Malayer University, 65719-95863 Malayer, Iran{}.}
\affiliation{$^{5}$Department of Physics, GC University, Lahore 54000, Pakistan.}
\affiliation{}
\date{\today }


\begin{abstract}
The low frequency, long wavelength electromagnetic waves, viz, shear Alfven
wave in quantum electron-positron-ion magneto plasmas, have been examined
using quantum magneto hydrodynamic model. In this model, we have considered
electrons and positrons are to be magnetized as well as degenerate whereas
ions are magnetized but classical. We have also included the effects of
exchange correlation terms which appear entirely the dynamic equations of
electrons and positrons. The whole treatment is done using multi-fluid model.
Our object is to study the shear Alfv\'{e}n waves propagating in above said
system of plasma. For that we have derived the modified dispersion relation of
the shear Alfv\'{e}n waves. Results are relevant to the terrestrial laboratory
astrophysics.

For correspondence: ehsan.zahida@gmail.com

\end{abstract}
\maketitle

\section{Introduction}

At about $10^{-6}<t<10$ $\sec$ time after the big bang when the universe was
evolving lepton epochs is believed to exist. In this particular epoch
temperatures of $10^{9}<T<10^{13}$ K are speculated to causing the
annihilation of the hadrons and antihadrons pairs which formed matter composed
of the electrons, positrons, and photons in thermodynamic equilibrium
\cite{1}. Pair plasmas consist of electrons and positrons (EP) inherent unique
properties due to their mass and charge symmetry and are known to exist
abundantly when early stars formation was taking place\cite{2,3,4,5,6,7}.
Their omnipresence is also accepted in the interior of accretion disks
surrounding black holes, magnetospheres of the neutron star and pulsar,
environments like the bipolar outflows (jets), active galactic nuclei, polar
regions of neutron stars, at the center of Milky Way galaxy etc.\cite{8,9,10}
In order to understand the physical phenomena happening hundred of light years
away, EP plasmas are created by interaction of ultraintense laser with the
solid targets. Scientists are hopeful that these interactions will lead to
creation of highly dense laboratory electron positron plasmas ($n\sim
10^{26}m^{-3}$) \cite{11,12}. Therefore theoretical investigations involving
pair plasma is also needed for the advancement of laboratory astrophysics.

Presence of ions, in EP plasmas has also been identified in both laboratory
and astrophysics\cite{13}, which breaks the symmetry of equal mass and number
of pair particles eventually leading to the new and interesting avenue of
research for the scientists\cite{9,10}.

Whereas Small temperature differences and also some nonlinear phenomena which
emerge naturally during the evolution of pair particles may usually cause this
asymmetric behavior in the experiments. On the other hand small contamination
of much heavier immobile ion, or small mass difference of the pair particles
can also produce asymmetries \cite{9,10}. In electron - positron - ion (EPI)
plasmas, physical phenomena like waves and instabilities can occur at both
fast (high frequency) and slow (low frequency) time scales. Research has been
carried out to study the both relativistic and non-relativistic pair plasmas
astrophysical nature and produced laboratory\cite{9, 13,14,15}. While in the
environment of neutron stars pair plasmas are speculated to be highly
degenerate and ultradense that is why a rigorous investigation for example in
the frame work of quantum hydrodynamics of EPI degenerate plasma has been made
over the past few years \cite{9}.

While shear Alfv\'{e}n wave features of magnetized plasmas are considered one
of the important waves in plasmas due to it's wide applications in both lab
and astrophysical environments. For their highly speculated importance
Alfv\'{e}n waves' propagation in electron positron plasmas with and without
ions have been extensively studied . For example since these waves damp much
slowly than the Langmuir or magneto-acoustic waves so are thought to cause
emission of electromagnetic radiations from the magnetosphere of pulsars
\cite{14,16,17}. 

Cerenkov radiation interaction with the plasma particles are speculated to as
the reason for the excitation shear Alfev waves however still there are
discrepancies and unexplainable features which require further
investigation\cite{18}. on the other hand as despite extensive theoretical
modelling, our knowledge of pair plamsas is still speculative, owing to the
extreme difficulty in recreating neutral matter-antimatter plasmas in the laboratory.

Since the reported possible creation of dense electron-positron plasma, where
the charged particles behave as a Fermi gas, and quantum mechanical effects
might play a vital role in the dynamics of charge carriers. It is important to
mention that the criterion of quantum interference of particles is satisfied
by the lighter plasma particles (electrons and positrons) more easily alos the
inclusion of the exchange-correlation potential with the quantum effects
through the Bohmian force and the quantum statistical pressure may reflect the
comprehensive study of a quantum plasma system \cite{19,20,21,22,23,24,25,26}.
The electron exchange and correlation effects in dense plasmas (by the
electron half-spin particles) play a central role in the plasma dielectric
response function \cite{27}. Hence, it is highly expected that contribution of
the electron-exchange potential along with the Bohm potential and the Fermi
degenerate pressure would reshape the dispersion properties of Alfv\'{e}n
waves and the interaction potentials of the medium in quantum plasmas.
Needless to mention electron exchange-correlation effects being inadequate
have been paid lass attention whereas for dense plasmas systems with low
temperature they can have dominant significance. 

The influence of quantum statistical degeneracy pressure and exchange
correlation effects on Shear Alfv\'{e}n waves in degenerate Fermi-Dirac
electron-positron ion plasma has not been investigated yet. Moreover since
pure Alfv\'{e}n waves propagate parallel to the magnetic field, are not
effected in quantized plasmas, whereas low frequency shear Alfv\'{e}n waves
making a small angle can be influenced by quantum effects so quite different
results than the classical Maxwellian plasmas can be expected. 

Therefore we aim to model for the dispersion of shear Alfv\'{e}n waves (SAWs)
in non-relativistic dense pair-ion plasmas with exchange correlation effects
which are attributes of  electrons and positrons only while doing so the
processes leading to the pair creation and recombination have been ignored. 

The manuscript is organized as follows: basic equations and the dispersion
relations for the shear Alfv\'{e}n waves propagating in electron-positron ions
are presented in section II. Quantitative analysis and conclusions are given
in Sec. III and Sec IV.

\section{\textbf{Basic Formulation and Instability Analysis}}

To study the dispersion properties of shear Alfv\'{e}n wave (SAW) making a
small angle with static external magnetic field $\mathbf{B}_{0}=B_{0}\hat{z}$
\ and propagating in a degenerate Fermi- Dirac pair-ion plasma, we write
quantum Euler equations for the $j$ species of electron positron and ion in
quantum Fermi-Dirac plasmas \cite{24}
\begin{equation}
\frac{\partial\mathbf{v}_{j}}{\partial t}=\frac{q_{j}}{m_{j}}\mathbf{E}%
+\mathbf{v}_{j}\times\omega_{cj}\hat{z}-\frac{\mathbf{\nabla}P_{Fj}}%
{m_{j}n_{0j}}+\frac{\hbar^{2}}{4m_{j}^{2}n_{0j}}\mathbf{\nabla}(\nabla
^{2}n_{j1})-V_{j,xc}\mathbf{\nabla}n_{j1}\label{1}%
\end{equation}

In above equation last term represents electron and positron
exchange-correlation potential which is a complex function of Fermi particles
density and is given as  $V_{j,xc}=\frac{0.985e^{2}}{\epsilon}n_{j}%
^{1/3}\left[  1+\frac{0.034}{a_{Bj}n_{j}^{1/3}}\ln\left(  1+18.37a_{Bj}%
n_{j}^{1/3}\right)  \right]  $\cite{28} is considered the attribute of the
spin effects in dense systems. For the readers it is useful to find that for
the degenerate plasma, these affects have been calculated comprehensibly in
\textquotedblleft Statistical Physics\textquotedblright\ book by Landau and
Lifshitz\cite{29} while exchange correlations for proton interaction have been
presented by Tsintsadze et al., \cite{30}. Since this depends upon the number
density, so we cannot ignore it in dense plasma environments. In Eq. (1)
$a_{Bj}=$\ $\epsilon\hbar^{2}/m_{j}e^{2}$ is the well-known Bohr atomic
radius. 

Equation (1) is general and conveniently written however later we will treat
ions as classical particle. In equation (\ref{1}) $\hbar=h/2\pi$ and
$\omega_{cj}=q_{j}B_{0}/m_{j}c$ the cyclotron frequency, $q_{j}$ the charge,
$m_{j}$ mass and $c$ is the velocity of light in a vacuum of the $j$th
species. Here, $j=i$ (ion), $j=e$ (electron), $j=p$ (positron), $q_{e}=-e$,
$q_{p}=+e$ and $q_{i}=Z_{i}e$, with $e$ being the magnitude of electronic
charge and $Z_{i}$ is the number of charges on ions. In Eq.(\ref{1}),
$P_{Fj}=\frac{m_{j}v_{Fj}^{2}n_{j}^{3}}{3n_{0j}^{2}}$ is pressure law for
3-dimensional Fermi gas \cite{28}, where $v_{Fj}^{2}=\frac{6}{5}\frac
{k_{B}T_{Fj}}{m_{j}}$ is the Fermi speed; $k_{B}$ is the Boltzmann constant,
$T_{Fj}=\frac{\hbar^{2}\left(  3\pi^{2}n_{0j}\right)  ^{2/3}}{2m_{j}}$\ is
Fermi temperature, $n_{j}=n_{0j}+n_{1j}$ the total number density with
equilibrium number density $n_{0j}$\ and perturbed number density $n_{1j}$\ of
$jth$ particles in the field of SAW. \ The ion component can be considered
classical or quantum depending upon the relevant parameters. However, in most
of the situations, ions are considered as cold fluid while describing the ion
wave. In these dense quantum and semiclassical plasmas, the screened
interaction potential cannot be characterized by the standard Debye-Huckel
model according to the multiparticle correlations and the quantum-mechanical
effects such as the Bohm potential, quantum pressure, and electron exchange
terms since the average kinetic energy of the plasma particle in quantum
plasmas is of the order of the Fermi energy \cite{28}. Thermal temperature of
ions is small as compared to the electrons and positrons and therefore ignored.

We assume the geometry of the problem that the SAW is propagating with low
frequency $\omega$\ on ion dynamics and obliquely to external magnetic filed
$\mathbf{B}_{0}=B_{0}\hat{z}$\ and the propagation vector $k$ of the wave
makes a small angle $\theta$\ with z - direction and lies in xz - plane i.e.,
$(k_{x},0,k_{z})$ where $k_{x}=k\sin\theta$ and $k_{z}=k\cos\theta$. The
velocity components of $j$th species in the field of SAW can be obtained from
Eq.(\ref{1}),%
\begin{equation}
v_{jx}=\frac{iq_{j}\left(  \omega^{2}F_{j}E_{x}+i\omega\omega_{cj}F_{j}%
E_{y}+V_{FBxcj}^{2}k_{x}k_{z}E_{z}\right)  }{m_{j}\omega\left(  \omega
^{2}-V_{FBxcj}^{2}k^{2}-F_{j}\omega_{cj}^{2}\right)  },\label{2}%
\end{equation}%
\begin{equation}
v_{jy}=\frac{iq_{j}\left[  -i\omega\omega_{cj}F_{j}E_{x}+\left(  \omega
^{2}-V_{FBxcj}^{2}k^{2}\right)  E_{y}-i\frac{\omega_{cj}V_{FBxcj}^{2}%
k_{x}k_{z}}{\omega}E_{z}\right]  }{m_{j}\omega\left(  \omega^{2}-V_{FBxcj}%
^{2}k^{2}-F_{j}\omega_{cj}^{2}\right)  }\label{3}%
\end{equation}
and%
\begin{equation}
v_{jz}=\frac{iq_{j}}{m_{j}\omega F_{j}}\left[
\begin{array}
[c]{c}%
\frac{V_{FBxcj}^{2}k_{x}k_{z}F_{j}}{\omega^{2}-V_{FBxcj}^{2}k^{2}-F_{j}%
\omega_{cj}^{2}}E_{x}+\frac{i\omega_{cj}V_{FBxcj}^{2}k_{x}k_{z}F_{j}}%
{\omega\left(  \omega^{2}-V_{FBxcj}^{2}k^{2}-F_{j}\omega_{cj}^{2}\right)
}E_{y}\\
+\left(  1+\frac{V_{FBxcj}^{4}k_{x}^{2}k_{z}^{2}}{\omega^{2}\left(  \omega
^{2}-V_{FBxcj}^{2}k^{2}-F_{j}\omega_{cj}^{2}\right)  }\right)  E_{z}%
\end{array}
\right]  ,\label{4}%
\end{equation}
where%
\begin{equation}
V_{FBxcj}^{2}=v_{Fj}^{2}+v_{Bj}^{2}+v_{xcj}^{2},\label{5}%
\end{equation}%
\begin{equation}
v_{Bj}^{2}=\frac{\hbar^{2}k^{2}}{4m_{j}^{2}},\label{6}%
\end{equation}%
\begin{equation}
v_{xcj}^{2}=\frac{0.985n_{0j}^{1/3}}{3m_{j}}\left(  \frac{e^{2}}{\epsilon
}\right)  \left[  1+\frac{0.034\times18.37}{1+18.37a_{Bj}n_{0j}^{1/3}}\right]
\label{7}%
\end{equation}
and%
\begin{equation}
F_{j}=1-\frac{V_{FBxcj}^{2}k_{z}^{2}}{\omega^{2}}.\label{8}%
\end{equation}
The electric field $E$ and magnetic field $B$ of SAW in pair ion plasma are
related by the following curl equations,
\begin{equation}
\mathbf{\nabla}\times\mathbf{{B}=}\frac{4\pi}{c}\mathbf{J+}\frac{1}{c}%
\frac{\partial\mathbf{E}}{\partial t}\mathbf{,}\label{9}%
\end{equation}
and
\begin{equation}
\mathbf{\nabla}\times\mathbf{{E}=-}\frac{1}{c}\frac{\partial\mathbf{B}%
}{\partial t}\mathbf{.}\label{10}%
\end{equation}
where
\begin{equation}
\mathbf{J}=\sum_{j}q_{j}n_{0j}\mathbf{v}_{j}\label{11}%
\end{equation}
is the current density of the plasma particles due to the propagation of
electromagnetic shear Alfv\'{e}n wave.

After substitution of Eqs.(\ref{2}-\ref{4}) into Eq.(\ref{11}), the current
density becomes,
\begin{equation}
\mathbf{J}=\underline{\underline{\mathbf{\sigma}}}.\mathbf{E}\label{12}%
\end{equation}
Where $\underline{\underline{\mathbf{\sigma}}}$ is the linear conductivity
tensor given by
\begin{equation}
\mathbf{\sigma}=\sum_{j}\frac{iq_{j}^{2}n_{0j}}{m_{j}\omega}\mathbf{K}%
_{j}\label{13}%
\end{equation}
where
\begin{equation}
\mathbf{K}_{j}=\left(
\begin{array}
[c]{ccc}%
\frac{\omega^{2}F_{j}}{\omega^{2}-V_{FBxcj}^{2}k^{2}-\omega_{cj}^{2}F_{j}} &
\frac{i\omega\omega_{cj}F_{j}}{\omega^{2}-V_{FBxcj}^{2}k^{2}-\omega_{cj}%
^{2}F_{j}} & \frac{V_{FBxcj}^{2}k_{x}k_{z}}{\omega^{2}-V_{FBxcj}^{2}%
k^{2}-\omega_{cj}^{2}F_{j}}\\
&  & \\
-\frac{i\omega\omega_{cj}F_{j}}{\omega^{2}-V_{FBxcj}^{2}k^{2}-\omega_{cj}%
^{2}F_{j}} & \frac{\omega^{2}-V_{FBxcj}^{2}k^{2}}{\omega^{2}-V_{FBxcj}%
^{2}k^{2}-\omega_{cj}^{2}F_{j}} & -i\frac{\omega_{cj}V_{FBxcj}^{2}k_{x}k_{z}%
}{\omega\left(  \omega^{2}-V_{FBxcj}^{2}k^{2}-\omega_{cj}^{2}F_{j}\right)  }\\
&  & \\
\frac{V_{FBxcj}^{2}k_{x}k_{z}}{\omega^{2}-V_{FBxcj}^{2}k^{2}-\omega_{cj}%
^{2}F_{j}} & i\frac{\omega_{cj}V_{FBxcj}^{2}k_{x}k_{z}}{\omega\left(
\omega^{2}-V_{FBxcj}^{2}k^{2}-\omega_{cj}^{2}F_{j}\right)  } & \frac{1}{F_{j}%
}\left(  1+\frac{V_{FBxcj}^{2}k_{x}^{2}k_{z}^{2}}{\omega^{2}\left(  \omega
^{2}-V_{FBxcj}^{2}k^{2}-\omega_{cj}^{2}F_{j}\right)  }\right)
\end{array}
\right)  .\label{14}%
\end{equation}
Combining these curl equations, we may write
\begin{equation}
\underline{\underline{\mathbf{D}}}.\mathbf{E}=0.\label{15}%
\end{equation}
$\underline{\underline{\mathbf{D}}}$ gives the linear plasma dispersion
relation due to electromagnetic shear Alfv\'{e}n wave $(\omega,k)$ and is
defined by
\begin{equation}
Det[\underline{\underline{\mathbf{D}}}]=k^{2}\underline{\underline{I}%
}-\mathbf{k}\mathbf{k}-\frac{\omega^{2}}{c^{2}}\underline{\underline{\epsilon
}}=0\label{16}%
\end{equation}
where $\underline{\underline{I}}$ is the unit dyadic and $\underline
{\underline{\epsilon}}=\underline{\underline{I}}-\sum_{j}\left(  \frac
{\omega_{pj}^{2}}{\omega^{2}}\right)  \underline{\underline{\mathbf{K}}}_{j}$.
Here, $\omega_{pj}=\left(  \frac{4\pi n_{0j}q_{j}^{2}}{m_{j}}\right)  ^{1/2}$
is the plasmas frequency of $j$th species. The matrix form of Eq.(\ref{16})
is
\begin{equation}
Det[\mathbf{D}]=Det\left(
\begin{array}
[c]{ccc}%
k_{z}^{2}-\frac{\omega^{2}}{c^{2}}\epsilon_{xx} & -\frac{\omega^{2}}{c^{2}%
}\epsilon_{xy} & -k_{z}k_{x}-\frac{\omega^{2}}{c^{2}}\epsilon_{xz}\\
&  & \\
-\frac{\omega^{2}}{c^{2}}\epsilon_{yx} & k^{2}-\frac{\omega^{2}}{c^{2}%
}\epsilon_{yy} & -\frac{\omega^{2}}{c^{2}}\epsilon_{yz}\\
&  & \\
-k_{z}k_{x}-\frac{\omega^{2}}{c^{2}}\epsilon_{zx} & -\frac{\omega^{2}}{c^{2}%
}\epsilon_{zy} & k_{x}^{2}-\frac{\omega^{2}}{c^{2}}\epsilon_{zz}%
\end{array}
\right)  =0.\label{17}%
\end{equation}
Here, we treat the electrons, positrons quantized and magnetized while the
ions are non-quantum but magnetized. The components of the medium response
function are,
\begin{equation}
\epsilon_{xx}=1-\frac{\omega_{pe}^{2}F_{e}}{\left(  \omega^{2}-V_{FBxce}%
^{2}k^{2}-\omega_{ce}^{2}F_{e}\right)  }-\frac{\omega_{pp}^{2}F_{p}}{\left(
\omega^{2}-V_{FBxcp}^{2}k^{2}-\omega_{cp}^{2}F_{p}\right)  }-\frac{\omega
_{pi}^{2}}{\omega^{2}-\omega_{ci}^{2}},\label{18}%
\end{equation}%
\begin{equation}
\epsilon_{yy}=1-\frac{\omega_{pe}^{2}\left(  1-V_{FBxce}^{2}k^{2}/\omega
^{2}\right)  }{\left(  \omega^{2}-V_{FBxce}^{2}k^{2}-\omega_{ce}^{2}%
F_{e}\right)  }-\frac{\omega_{pp}^{2}\left(  1-V_{FBxcp}^{2}k^{2}/\omega
^{2}\right)  }{\left(  \omega^{2}-V_{FBxcp}^{2}k^{2}-\omega_{cp}^{2}%
F_{p}\right)  }-\frac{\omega_{pi}^{2}}{\omega^{2}-\omega_{ci}^{2}},\label{19}%
\end{equation}%
\begin{equation}%
\begin{array}
[c]{c}%
\epsilon_{zz}=1-\frac{\omega_{pe}^{2}}{\omega^{2}F_{e}}\left(  1+\frac
{V_{FBxce}^{4}k_{x}^{2}k_{z}^{2}}{\omega^{2}\left(  \omega^{2}-V_{FBxce}%
^{2}k^{2}-\omega_{ce}^{2}F_{e}\right)  }\right)  \\
-\frac{\omega_{pp}^{2}}{\omega^{2}F_{p}}\left(  1+\frac{V_{FBxcp}^{4}k_{x}%
^{2}k_{z}^{2}}{\omega^{2}\left(  \omega^{2}-V_{FBxcp}^{2}A_{p}^{2}k^{2}%
-\omega_{cp}^{2}F_{p}\right)  }\right)  -\frac{\omega_{pi}^{2}}{\omega^{2}},
\end{array}
\label{20a}%
\end{equation}%
\begin{equation}
\epsilon_{xy}=i\frac{\omega_{pe}^{2}F_{e}}{\left(  \omega^{2}-V_{FBxce}%
^{2}k^{2}-\omega_{ce}^{2}F_{e}\right)  }\frac{\omega_{ce}}{\omega}%
-i\frac{\omega_{pp}^{2}F_{p}}{\left(  \omega^{2}-V_{FBxcp}^{2}k^{2}%
-\omega_{cp}^{2}F_{p}\right)  }\frac{\omega_{cp}}{\omega}-i\frac{\omega
_{pi}^{2}}{\omega^{2}-\omega_{ci}^{2}}\frac{\omega_{ci}}{\omega},\label{21}%
\end{equation}%
\begin{equation}
\epsilon_{xz}=-\left(  \frac{\omega_{pe}^{2}V_{FBxce}^{2}k_{x}k_{z}}%
{\omega^{2}\left(  \omega^{2}-V_{FBxce}^{2}k^{2}-\omega_{ce}^{2}F_{e}\right)
}\right)  -\left(  \frac{\omega_{pp}^{2}V_{FBxcp}^{2}k_{x}k_{z}}{\omega
^{2}\left(  \omega^{2}-V_{FBxcp}^{2}k^{2}-\omega_{cp}^{2}F_{p}\right)
}\right)  ,\label{22}%
\end{equation}%
\begin{equation}
\epsilon_{yz}=-i\left(  \frac{\omega_{ce}\omega_{pe}^{2}V_{FBxce}^{2}%
k_{x}k_{z}}{\omega^{3}\left(  \omega^{2}-V_{FBxce}^{2}k^{2}-\omega_{ce}%
^{2}F_{e}\right)  }\right)  +i\left(  \frac{\omega_{cp}\omega_{pp}%
^{2}V_{FBxcp}^{2}k_{x}k_{z}}{\omega^{3}\left(  \omega^{2}-V_{FBxcp}^{2}%
k^{2}-\omega_{cp}^{2}F_{p}\right)  }\right)  ,\label{23}%
\end{equation}%
\begin{equation}
\epsilon_{yx}=-\epsilon_{xy},\text{ \ \ \ }\epsilon_{zx}=\epsilon_{xz},\text{
\ \ \ }\epsilon_{zy}=-\epsilon_{yz}.\label{24}%
\end{equation}
For the oblique SAW case the propagation vector and the electric field are
parallel to eachother and so contribution of y component of electric field can
ignored to zero and so Eq.(\ref{17}) can be reduced to %

\begin{equation}
Det[\underline{\underline{\mathbf{D}}}]=Det\left(
\begin{array}
[c]{cc}%
k_{z}^{2}-\frac{\omega^{2}}{c^{2}}\epsilon_{xx} & -k_{z}k_{x}-\frac{\omega
^{2}}{c^{2}}\epsilon_{xz}\\
& \\
-k_{z}k_{x}-\frac{\omega^{2}}{c^{2}}\epsilon_{zx} & k_{x}^{2}-\frac{\omega
^{2}}{c^{2}}\epsilon_{zz}%
\end{array}
\right)  =0.\label{25}%
\end{equation}
Or, the above equation can be written as follows,
\begin{equation}
\omega^{2}(\epsilon_{xx}\epsilon_{zz}-\epsilon_{xz}^{2})-c^{2}k_{z}%
^{2}\epsilon_{zz}-c^{2}k_{x}^{2}\epsilon_{xx}-2c^{2}k_{x}k_{z}\epsilon
_{xz}=0\label{26}%
\end{equation}
Firstly, the mass of electrons and positrons \ are ignored for being much
lighter than the heaviour ions. Then, for frequency range $\omega^{2}\ll
\omega_{ci}^{2}\ll\omega_{ce}^{2}=\omega_{cp}^{2}$ and $\omega^{2}\ll
V_{FBxce}^{2}k_{z}^{2}$, $\omega^{2}\ll V_{FBxcp}^{2}k_{z}^{2}$ the components
of medium response function gain following simplified form,
\begin{equation}
\epsilon_{xz}=-\frac{\omega_{pe}^{2}}{\omega_{ce}^{2}}\frac{k_{x}}{k_{z}%
}-\frac{\omega_{pp}^{2}}{\omega_{cp}^{2}}\frac{k_{x}}{k_{z}}\label{27}%
\end{equation}%
\begin{equation}
\epsilon_{xx}=1+\frac{\omega_{pe}^{2}}{\omega_{ce}^{2}}+\frac{\omega_{pp}^{2}%
}{\omega_{cp}^{2}}+\frac{\omega_{pi}^{2}}{\omega_{ci}^{2}},\label{28}%
\end{equation}%
\begin{equation}
\epsilon_{zz}=1+\frac{\omega_{pe}^{2}}{V_{FBxce}^{2}k_{z}^{2}}+\frac
{\omega_{pe}^{2}}{\omega_{ce}^{2}}\frac{k_{x}^{2}}{k_{z}^{2}}+\frac
{\omega_{pp}^{2}}{V_{FBxcp}^{2}k_{z}^{2}}+\frac{\omega_{pp}^{2}}{\omega
_{cp}^{2}}\frac{k_{x}^{2}}{k_{z}^{2}}-\frac{\omega_{pi}^{2}}{\omega^{2}%
},\label{29}%
\end{equation}
using Eqs.(\ref{27}-\ref{29}) into Eq.(\ref{26}) we get
\begin{equation}
A\omega^{4}+B\omega^{2}+C=0\label{30}%
\end{equation}
where%
\begin{equation}
A=\left\{  \left(  1+\frac{c^{2}}{v_{A}^{2}}\right)  \left(  1+\frac{1}%
{k_{z}^{2}\lambda_{DFc}^{2}}+\frac{c^{2}}{v_{Aep}^{2}}\frac{k_{x}^{2}}%
{k_{z}^{2}}\right)  -\frac{c^{4}}{v_{Aep}^{4}}\frac{k_{x}^{2}}{k_{z}^{2}%
}\right\}  \label{31}%
\end{equation}%
\begin{equation}
B=-\left\{  \left(  1+\frac{c^{2}}{v_{A}^{2}}\right)  \omega_{pi}^{2}%
+c^{2}k_{z}^{2}\left(  1+\frac{1}{k_{z}^{2}\lambda_{DFc}^{2}}\right)
+c^{2}k_{x}^{2}\left(  1+\frac{c^{2}}{v_{Ai}^{2}}\right)  \right\}  \label{32}%
\end{equation}
and
\begin{equation}
C=c^{2}k_{z}^{2}\omega_{pi}^{2}\label{33}%
\end{equation}
For simplifying $A,$ $B$ and $C,$ here we have used $v_{Ai}^{2}=c^{2}%
\frac{\omega_{ci}^{2}}{\omega_{pi}^{2}},v_{Ae}^{2}=c^{2}\frac{\omega_{ce}^{2}%
}{\omega_{pe}^{2}},v_{Ap}^{2}=c^{2}\frac{\omega_{cp}^{2}}{\omega_{pp}^{2}}%
,$\ $\frac{1}{v_{A}^{2}}=\left(  \frac{1}{v_{Ae}^{2}}+\frac{1}{v_{Ap}^{2}%
}+\frac{1}{v_{Ai}^{2}}\right)  $, $\frac{1}{v_{Aep}^{2}}=\left(  \frac
{1}{v_{Ae}^{2}}+\frac{1}{v_{Ap}^{2}}\right)  ,$ $\lambda_{DFe}^{2}%
=\frac{V_{FBxce}^{2}}{\omega_{pe}^{2}},$ $\lambda_{DFp}^{2}=\frac
{V_{FBxcp}^{2}}{\omega_{pp}^{2}},$ and $1/\lambda_{DFc}^{2}=\left(
1/\lambda_{DFe}^{2}+1/\lambda_{DFp}^{2}\right)  $. For further simplification,
using $\lambda_{DFc}^{2}k_{z}^{2}\ll1,$ $v_{Ai}^{2}\ll c^{2}$ in above
expressions of $A,$ $B$ and $C$ we get%

\begin{equation}
A=\left\{  \frac{c^{2}}{v_{A}^{2}}\frac{1}{k_{z}^{2}\lambda_{DFc}^{2}}%
+\frac{c^{4}}{v_{Ai}^{2}v_{Aep}^{2}}\frac{k_{x}^{2}}{k_{z}^{2}}\right\}
\label{34}%
\end{equation}%
\begin{equation}
B=-c^{2}\left\{  \frac{\omega_{pi}^{2}}{v_{A}^{2}}+\frac{1}{\lambda_{DFc}^{2}%
}+\frac{c^{2}k_{x}^{2}}{v_{Ai}^{2}}\right\}  \label{35}%
\end{equation}%
\begin{equation}
C=c^{2}k_{z}^{2}\omega_{pi}^{2}\label{36}%
\end{equation}
Eq.(\ref{30}) is quadratic in $\omega^{2}$ then%
\begin{equation}
\omega^{2}=\frac{\left(  \frac{\omega_{pi}^{2}}{v_{A}^{2}}+\frac{1}%
{\lambda_{DFc}^{2}}+\frac{c^{2}k_{x}^{2}}{v_{Ai}^{2}}\right)  \pm\sqrt{\left(
\frac{1}{\lambda_{DFc}^{2}}+\frac{c^{2}k_{x}^{2}}{v_{Ai}^{2}}-\frac
{\omega_{pi}^{2}}{v_{A}^{2}}\right)  ^{2}-4\omega_{pi}^{2}\frac{c^{2}k_{x}%
^{2}}{v_{Ai}^{4}}}}{2\left(  \frac{1}{v_{A}^{2}}\frac{1}{k_{z}^{2}%
\lambda_{DFc}^{2}}+\frac{c^{2}}{v_{Ai}^{2}v_{Aep}^{2}}\frac{k_{x}^{2}}%
{k_{z}^{2}}\right)  }\label{37}%
\end{equation}
for real frequency the term in square root should be positive this implies
that \ $\left(  \frac{1}{\lambda_{DFc}^{2}}+\frac{c^{2}k_{x}^{2}}{v_{Ai}^{2}%
}-\frac{\omega_{pi}^{2}}{v_{A}^{2}}\right)  ^{2}>>-4\omega_{pi}^{2}\frac
{c^{2}k_{x}^{2}}{v_{Ai}^{4}}$ then%
\begin{equation}
\omega^{2}=\frac{\left(  \frac{\omega_{pi}^{2}}{v_{A}^{2}}+\frac{1}%
{\lambda_{DFc}^{2}}+\frac{c^{2}k_{x}^{2}}{v_{Ai}^{2}}\right)  \pm\left(
\frac{1}{\lambda_{DFc}^{2}}+\frac{c^{2}k_{x}^{2}}{v_{Ai}^{2}}-\frac
{\omega_{pi}^{2}}{v_{A}^{2}}\right)  }{2\left(  \frac{1}{v_{A}^{2}}\frac
{1}{k_{z}^{2}\lambda_{DFc}^{2}}+\frac{c^{2}}{v_{Ai}^{2}v_{Aep}^{2}}\frac
{k_{x}^{2}}{k_{z}^{2}}\right)  }\label{38}%
\end{equation}
for the propagation of SAW using upper sign term and after some simplification
we get%

\begin{equation}
\omega^{2}=\frac{k_{z}^{2}v_{A}^{2}\left(  1+k_{x}^{2}\rho_{Fi}^{2}\right)
}{1+v_{A}^{2}/v_{Aep}^{2}k_{x}^{2}\rho_{Fi}^{2}}\label{39}%
\end{equation}
where $\rho_{Fi}^{2}=\left(  c^{2}/v_{Ai}^{2}\right)  \lambda_{DFc}^{2}.$ This
is the dispersion relation of SAW in electron positron-ion plasmas modified by
exchange-correlation potentials due to electrons and positrons which is
modified from the one in ordinary classical plasmas\cite{31}.

\section{Results and Discussion}

Electron-positron ion plasma exist in very dense astrophysical environment
with electron number density $n_{0e}\sim10^{27}cm^{-3}$ and in laboratory with
$n_{0e}\sim10^{16}cm^{-3}$. Now we quantitatively analyse the results obtained
in Sec. (2). In this study, the typical parameters for dense plasmas \cite{2}
that are relevant to astrophysical objects e.g., neutron stars and pulsar's
atmosphere have been used. In such environments, due to highly density and
strong magnetic field of many orders higher than that of laboratory plasma,
behaves exotically. The interaction between positrons and/or electrons in such
plasmas is very weak due to Pauli blocking and they are more suited for
quantum hydrodynamics. Therefore, we select the following typical electron
number density $n_{0e}\sim1.5\times10^{22}cm^{-3},$ with very high external
magnetic field $B_{0}\sim10^{10}G$ and using physical constants in cgs system
viz., $c=3\times10^{10}cm\sec^{-1},$ $m_{e}=9.1\times10^{-28}g,$
$m_{i}=1.67\times10^{-24}g$, and $\hbar=1.057\times10^{-27}erg\sec$. Eq.
(\ref{39}) is plotted to investigate the changes in the dispersion
characteristics of shear Alfv\'{e}n waves for different variables that are:
($\omega/\omega_{ci}$ vs $k\frac{v_{A}}{\omega_{ci}})$, ($\omega/\omega_{ci}$
vs $n_{0i}[cm^{-3}]),$ ($\omega/\omega_{ci}$ vs $B_{0}[G])$ and ($\omega
/\omega_{ci}$ vs $\theta\lbrack Degree]).$

Figure (1) shows the plot of Eq. (39) and increase in frequency $\omega$ and
phase speed of SAW with propagation vector $k$ can be noticed. It represents
that for any value of $k,$ the frequency $\omega$ and phase speed\ of SAW
increases due the inclusion of exchange - correlation Potential. The plot is a
curve due to the oblique propagation of SAW with external magnetic field (see
equation (\ref{39})). Thus oblique propagation of SAW effected by Fermi
temperature and Exchange potential. For Alfv\'{e}n wave propagating exactly
along ambient magnetic field that is $k_{x}=0$ then we left with the
dispersion relation $\omega^{2}=k_{z}^{2}v_{A}^{2},$ which cannot be effected
by the contribution of Bohm, Fermi and Exchange potential.

Fig.(2) shows the increment in the frequency band with the increase of ion
number density. It also shows the growth of frequency with the consideration
of Exchange-Potential. Fig.(3) explains that the Alfv\'{e}n frequency
decreases with strength of magnetic field as expected. But there are larger
values of $\omega$ at every value of magnetic field with exchange potential
than values of $\omega$ without exchange-potential. In Fig.(4) there is a
large difference in the frequency of wave with and without exchange-potential
on the increment of small value of $\theta.$

Concluding, dispersion relation of low frequency electromagnetic waves or
shear Alfv\'{e}n wave nonrelativistic Fermi-Dirac pair ion plasma have been
derived and studied quantitatively. Overall the exchange correlation affects
significantly modify the waves depression. These results are in particularly
important for the terrestrial laboratory astrophysics as future ultra-intense
lasers are believed to produce dense degenerate pair plasmas under favorable
conditions.

One of us (Z.E.) is grateful to Professor Nodar Tsintsadze for the discussions
which helped to understand this problem better. M. S is thankful to A. Rasheed
for fruitful discussions. Financial support from higher education commission
of Pakistan through grant No.2323/NRPU 

and No.21-479/SRGP/R\&D/HEC/2014 are highly appriciated.

{\LARGE Figu}{\Large re Captions}

Fig. 1: Relationship of [$\omega/\omega_{ci}$ vs $k\frac{v_{A}}{\omega_{ci}}]$
with Exchange-Potential (solid curve) and without Exchange-Potential (dashed
curve) at $n_{0e}\sim1.5\times10^{22}cm^{-3},$ $B_{0}\sim10^{10}G,$
$\theta\sim5[Degree].$

Fig. 2: Relationship of [$\omega/\omega_{ci}$ vs $n_{0i}[cm^{-3}]]$ with
Exchange-Potential (solid curve) and without Exchange-Potential (dashed curve)
at $\theta\sim5[Degree],$ $B_{0}\sim10^{10}G.$

Fig. 3: Relationship of [$\omega/\omega_{ci}$ vs $B_{0}[G]]$ with
Exchange-Potential (solid curve) and without Exchange-Potential (dashed curve)
at $n_{0e}\sim1.5\times10^{22}cm^{-3},$ $\theta\sim5[Degree].$

Fig. 4: Relationship of [$\omega/\omega_{ci}$ vs $\theta\lbrack Degree]]$ with
Exchange-Potential (solid curve) and without Exchange-Potential (dashed curve)
at $n_{0e}\sim1.5\times10^{22}cm^{-3},$ $B_{0}\sim10^{10}G.$
\end{document}